\documentclass[preprint,showpacs,aps,pra]{revtex4-1}
\usepackage{epsfig}
\usepackage{psfrag}
\usepackage{graphicx}
\usepackage{titlesec}
\usepackage{graphics}
\usepackage{amsmath}
\usepackage{color}
\usepackage[title]{appendix}

\begin{document}


\title{Modified plasma waves described by a logarithmic electrodynamics}

\author{Fernando Haas} \email{fernando.haas@ufrgs.br}
\affiliation{Instituto de F\'{\i}sica, Universidade Federal do Rio Grande do Sul, Av. Bento Gon\c{c}alves 9500, 91501-970 Porto Alegre, RS, Brasil}

\author{Patricio Gaete} \email{patricio.gaete@usm.cl} 
\affiliation{Departamento de F\'{i}sica and Centro Cient\'{i}fico-Tecnol\'ogico de Valpara\'{i}so-CCTVal,
Universidad T\'{e}cnica Federico Santa Mar\'{i}a, Valpara\'{i}so, Chile}

\author{L.P.R. Ospedal} \email{leoopr@cbpf.br}
\affiliation{Centro Brasileiro de Pesquisas F\'{i}sicas (CBPF), Rio de Janeiro, RJ, Brasil} 

\author{J.A. Helay\"{e}l-Neto}\email{helayel@cbpf.br}
\affiliation{Centro Brasileiro de Pesquisas F\'{i}sicas (CBPF), Rio de Janeiro, RJ, Brasil}

\begin{abstract}

The propagation of plasma waves in a new non-linear, logarithmic electrodynamics model is performed. A cold, uniform, collisionless fluid plasma model is applied.  Electrostatic waves in magnetized plasma are shown to correspond to modified Trivelpiece-Gould modes, together with changes of the plasma and upper-hybrid frequencies, driven by the logarithmic electrodynamics effects. Electromagnetic waves are described by a generalized Appleton-Hartree dispersion relation. The cases of propagation parallel or perpendicular to the equilibrium magnetic field are analyzed in detail. In particular, generalized ordinary and extraordinary modes are obtained. We determine the changes, due to logarithmic electrodynamics, in the allowable and forbidden frequency bands of the new extraordinary mode. Estimates are provided about the strength of the ambient magnetic field, so that the non-linear electrodynamics effects become decisive.

\end{abstract}


\maketitle

\section{Introduction}

In $1923$, the experiments by Compton on the scattering of $X-$rays off electrons demonstrated that Einstein's light quanta carry not only energy, but also momentum, making it clear that they behave as true particles \cite{a}. It was Lewis, in his $1926$ article, {\it The Conservation of Photons} \cite{b}, who coined the word photon to name the particles of light, though, to his sense, the photon was understood as a sort of atom of light. One year later, in the $1927$ Solvay Meeting, entitled {\it Electrons and Photons}, Compton used the term photon as we understand it today \cite{c}. 

Ever since, there started a broad activity aimed to study the light by light scattering, once the light quanta $-$ the photons $-$ were then understood as genuine elementary particles and, therefore, they could scatter each other. In $1930$ and $1931$, a series of papers was devoted to detect the collision between photons and to check whether or not the superposition principle was respected \cite{d}. The perception was that the photon-photon scattering could unveil non-linear effects in the electromagnetic theory once deviations from the superposition principle were detected. However, only in $1933$, a theoretical investigation, based on the $1928$ Dirac's theory for the electron \cite{e}, was proposed by Otto Halpern \cite{f}, who claimed that virtual electron-positron pairs could be the actual origin of photon-photon collisions. This short $-$ but deep and consequent $-$ work provided a more qualitative framework to be applied in the description of the photon-photon scattering. Halpern's paper opened up a very intensive line of investigation in the immediate following years. Later on, in $1935$, Euler and Kockel, both Heisenberg's students, based on the early developments of Quantum Electrodynamics (QED), derived the leading non-linear corrections to the Maxwell equations in vacuum \cite{g}. The years $1933-1936$ were a rich period for the inspection of non-linear electrodynamic models, when the Delbr\"{u}ck scattering \cite{h}, the Breit-Wheeler effect \cite{i}, the spontaneous decay of photons \cite{j} and the Euler-Heisenberg \cite{k} and Born-Infeld \cite{l} models were investigated. Photon-photon scattering and the detection of physical phenomena in vacuum as a consequence of non-linear electrodynamics (NLED) remain a challenge up to present days for both theoretical and experimental physics \cite{MM}.

More recently, there has been a renewed interest in NLED in connection with new Physics beyond the Standard Model, specially the possibility of detecting non-linear electromagnetic vacuum effects induced by quantum gravity corrections to Maxwell electrodynamics \cite{m}. In connection with black holes, one has focused on different NLED models to get a broad class of singularity-free black hole solutions \cite{n}. In a recent paper \cite{o}, a logarithmic electrodynamic action has been presented and inspected to analyze the thermodynamic implications of NLED on an $AdS$ black hole solution. The present paper sets out to pursue an investigation of this particular (logarithmic) NLED in the context of a plasma, as it shall be described in the sequel.

The propagation of electromagnetic waves in a logarithmic electrodynamics context has already been derived in free space \cite{arXiv}. However, the analysis of wave propagation in a material medium governed by logarithmic electrodynamics has not yet been carried out. In a first approach to this subject, the present work considers a cold, non-relativistic, collisionless plasma, composed by electrons inn a fixed homogeneous ionic background. The choice of the simplest possible plasma allows to investigate the main new effects induced by logarithmic electrodynamics on some of the most salient plasma waves. It would be impossible to perform a similar analysis for all relevant plasma waves. For instance, in this first attempt, magnetohydrodynamic waves or waves taking into account the ions response (e.g. ion-acoustic waves) are not yet addressed. Similarly, high amplitude, non-linear waves are not considered here. Nevertheless, a rich variety of logarithmic- electrodynamics-driven essential new aspects of wave propagation in plasmas shall be identified. A basic result common to all classes of waves considered in our analysis is that logarithmic electrodynamic effects crucially depend on the parameter $c {\bf B}_0/\beta$, where $c$ is the speed of light, ${\bf B}_0$ is the equilibrium magnetic field and $\beta$ is a fundamental parameter of the model, which we presently consider to be positive. The general result can be of help for the determination of $\beta$, which is here taken as a phenomenological parameter within logarithmic electrodynamics \cite{arXiv}. 

This work is organized according to the outline that follows. To go through a logical path, we start in Section II from the simplest plasma waves, viz. electrostatic waves, in both unmagnetized and magnetized plasma. In Section III,  electromagnetic perturbations are allowed, which are known to be described by the Appleton-Hartree equation \cite{AH}, for arbitrary propagation angle. Here, a modified Appleton-Hartree equation is obtained, with changes driven by the logarithmic electrodynamics. In Section IV, we treat the special case of waves propagating along the direction of the ambient magnetic field. In Section V, the perpendicular wave propagation is addressed to, along with the derivation of the corresponding modified ordinary and extraordinary modes. Section VI shows some physical estimates on the strength of the effects stemming from logarithmic electrodynamics. In Section , we cast our Concluding Comments.  

\section{Electrostatic waves}

According to \cite{arXiv}, logarithmic electrodynamics is described  by the following Lagrangian density
\begin{equation}
\mathcal{L} = - 2 \beta^2 \varepsilon_0 \, \ln \left[ 1 + \frac{1}{\beta} \, \sqrt{ \frac{s}{2} \, \biggl( c^2 |{\bf B}|^2 - |{\bf E}|^2 \biggr) } \, \right]   
+ 2 \beta \, \varepsilon_0 \, \sqrt{ \frac{s}{2} \, \biggl( c^2 |{\bf B}|^2 - |{\bf E}|^2 \biggr) } \, ,
\label{L-log} \end{equation}
where ${\bf E}$ and ${\bf B}$ denote the electric and magnetic fields, respectively. In addition, $\beta$ is a fundamental physical parameter, $c$ is the speed of light and $s$ is a number such that $s = -1$ for $|{\bf E}| \geq c |{\bf B}|$ and $s = 1$ for $ c |{\bf B}| > |{\bf E}|$.

The corresponding field equations, in the presence of charge and current densities $\rho$ and ${\bf j}$, read as below:
\begin{eqnarray}
\nabla \cdot {\bf D} &=& \frac{\rho}{\varepsilon_0} \, , \label{ED_0} \\
\nabla \times {\bf E} &=& - \frac{\partial {\bf B}}{\partial t} \, , \label{ED_1} \\
\nabla \cdot {\bf B} &=& 0 \, , \label{ED_2} \\
\nabla \times {\bf H} &=& \mu_0 \, {\bf j} + \frac{1}{c^2} \frac{\partial {\bf D}}{\partial t} \, ,
\label{ED_3}
\end{eqnarray}
where $\varepsilon_0$ is the vacuum permittivity, $\mu_0$ the vacuum permeability and the auxiliary fields ${\bf D}$ and ${\bf H}$ are given by
\begin{equation}
{\bf D} = \frac{\sqrt{2}\beta{\bf E}}{\sqrt{2}\beta + \sqrt{-s(|{\bf E}|^2-c^2 |{\bf B}|^2)}} \,, \quad {\bf H} = \frac{\sqrt{2}\beta{\bf B}}{\sqrt{2}\beta + \sqrt{-s(|{\bf E}|^2-c^2 |{\bf B}|^2)}} \, . \label{x1}
\end{equation}

In a first approach, let us focus on the  electrostatic case, using fluid theory in the cold plasma limit, 
\begin{eqnarray}
\frac{\partial n}{\partial t} + \nabla\cdot(n{\bf u}) &=& 0 \,, \label{e24} \\
\frac{\partial{\bf u}}{\partial t} + {\bf u}\cdot\nabla{\bf u} &=& - \frac{e}{m}({\bf E} + {\bf u}\times{\bf B}_0) \,, \label{e25} \\
\nabla\cdot{\bf D} &=& \frac{e}{\varepsilon_0} (n_0 - n) \,. \label{e26} 
\end{eqnarray}
Here, $n$ is the electrons' number density, ${\bf u}$ is the electrons' fluid velocity field, $-e$ and $m$ are respectively the electron charge and mass, ${\bf B}_0$ is the equilibrium magnetic field and $n_0$ is the ions' background number density (with atomic number $Z = 1$). For simplicity, ions are supposed to be infinitely massive, which is appropriate for high frequency waves. Likewise, thermal or collisional effects are also disregarded. 

Linear electrostatic waves consider $n = n_0 + \delta n$, ${\bf u} = \delta{\bf u}$ and ${\bf E} = \delta{\bf E}$, where $\delta n, \delta{\bf u}$ and $\delta{\bf E}$ are first order plane wave perturbations proportional to $\exp[i({\bf k}\cdot{\bf r} - \omega t)]$, where ${\bf k}$ is the wave vector and $\omega$ is the (angular) wave frequency. Magnetic field perturbations shall start to be studied in Section III. Hence, at this stage, the ${\bf H}$ field is not required. In passing, we note that for a cold plasma and for linear waves with zero equilibrium fluid velocity, a non-relativistic treatment is sufficient.

Initially, in the unmagnetized case (${\bf B}_0 = 0$), we have, from Eq. (\ref{x1}), that $\delta{\bf D} = \delta{\bf E}$. Therefore, in this situation, we detect no changes due to the logarithmic electrodynamics, at least for small amplitude, linear waves. The usual (non-propagating) electron plasma wave with $\omega^2 = \omega_p^2$ is recovered, where $\omega_p = [n_0 e^2/(m \varepsilon_0)]^{1/2}$ is the plasma frequency. 

In the magnetized case, one has $|{\bf B}_0| = B_0 > |{\bf E}|/c$, since the magnetic field is finite and the electric field is a perturbation. Hence, from Eq. (\ref{x1}), we have $s = 1$ and 
\begin{equation}
{\bf D} = \frac{\sqrt{2}\beta{\bf E}}{\sqrt{2}\beta + \sqrt{c^2 B_0^2 - |{\bf E}|^2}} \,. \label{e27}
\end{equation}
Linearizing Eqs. (\ref{e24})-(\ref{e27}) and assuming ${\bf B} = B_0 \hat{z}, {\bf k} = k \hat{x}, {\bf E} = \delta E \hat{x}$ (longitudinal wave) yield 
\begin{equation}
\label{e28}
\omega^2 = \tilde{\omega}_h^2 = \tilde{\omega}_p^2 + \omega_c^2 \,, 
\end{equation}
where $\omega_c = e B_0/m$ is the electron's cyclotron frequency, 
\begin{equation}
\label{e29}
\tilde{\omega_p} = \omega_p \left(1 + \frac{c B_0}{\sqrt{2}\beta}\right)^{1/2}
\end{equation}
is a modified plasma frequency and $\tilde{\omega}_h$ is a modified upper-hybrid frequency.  In the limit $B_0 \ll \beta$, one re-obtains the usual upper hybrid wave $\omega^2 = \omega_h^2 = \omega_p^2 + \omega_c^2$. Otherwise, Eq. (\ref{e28}) shows a modified upper hybrid wave due to log effects. 

With more generality, propagation in an arbitrary direction so that ${\bf k} \parallel {\bf E}$ but with ${\bf k} = k (\sin\theta, 0 , \cos\theta)$ gives a modified Trivelpiece-Gould dispersion relation, 
\begin{equation}
\label{e30}
\omega^4 - (\tilde{\omega}_p^2 + \omega_c^2)\,\omega^2 + \tilde{\omega}_p^2\omega_c^2\cos^{2}\theta = 0 \,,
\end{equation}
yielding
\begin{equation}
\label{e31}
\omega^2 = \frac{1}{2}\left[\tilde{\omega}_p^2 + \omega_c^2 \pm \left((\tilde{\omega}_p^2 - \omega_c^2)^2 + 4 \tilde{\omega}_p^2 \omega_c^2 \sin^{2}\theta\right)^{1/2}\right] \,.
\end{equation}
For $\theta = \pi/2$, we recover Eq. (\ref{e28}). The modes following from Eq. (\ref{e31}) are always stable ($\omega^2 \geq 0$). In the low field case $c B_0 \ll \beta$, we recover the usual Trivelpiece-Gould modes. Notice that the original work by Trivelpiece and Gould \cite{TG} has considered wave propagation along an arbitrary angle, see also Fig. 4.21 in \cite{Chen}.

\section{Generalized Appleton-Hartree equation}

We now turn to electromagnetic electron waves in cold, collisionless magnetized plasma in a fixed homogeneous ionic background. The relevant equations are the Faraday's law, Eq. \eqref{ED_1}, the force equation, namely,
\begin{equation}
\frac{\partial{\bf u}}{\partial t} + {\bf u}\cdot\nabla{\bf u} = - \frac{e}{m}({\bf E} + {\bf u}\times{\bf B}) \,, \label{y1} \\
\end{equation}
and the modified Amp\`ere-Maxwell law, Eq. \eqref{ED_3}, which can be recast as 
\begin{equation}
\nabla\times{\bf H} = - \mu_0 n e {\bf u} + \frac{1}{c^2}\frac{\partial{\bf D}}{\partial t} \, . \label{y2}
\end{equation}

Following the usual procedure \cite{Stix, Swanson, Bittencourt}, we are going to linearize around the homogeneous equilibrium $n = n_0, {\bf u} = 0, {\bf E} = 0, {\bf B} = {\bf B_0} \neq 0$. In passing, notice that from Eq. (\ref{x1}) for ${\bf B}_0 = 0$ gives $\delta{\bf D} = \delta{\bf E}, \delta{\bf H} = \delta{\bf B}$, whatever the sign of $s$, so that no logarithmic effects show up in this case, which is associated to the standard electromagnetic plasma wave $\omega^2 = \omega_p^2 + c^2 k^2$ for ${\bf k}\cdot{\bf E} = 0$ and to the electron plasma mode $\omega^2 = \omega_p^2$ for ${\bf k}\cdot{\bf E} \neq 0$. 

As in the previous Section, in the magnetized case for linear waves, we must set $s = 1$. We have 
\begin{equation}
{\bf D} = \frac{\sqrt{2}\beta{\bf E}}{\sqrt{2}\beta + \sqrt{c^2 |{\bf B}|^2 - |{\bf E}|^2}} \,, \quad {\bf H} = \frac{\sqrt{2}\beta{\bf B}}{\sqrt{2}\beta + \sqrt{c^2 |{\bf B}|^2 - |{\bf E}|^2}} \,. \label{x27}
\end{equation}
Assuming $n = n_0 + \delta n, {\bf u} = \delta{\bf u}, {\bf E} = \delta{\bf E}, {\bf B} = {\bf B}_0 + \delta{\bf B}$, for small amplitude perturbation, yields, in particular,
\begin{equation}
\delta{\bf D} = \frac{H_0}{B_0}\delta{\bf E} \,, \quad {\bf H} = {\bf H}_0 + \delta{\bf H} \,, \label{y3}
\end{equation}
where $H_0 = |{\bf H}_0|$, 
\begin{equation}
{\bf H}_0 = \frac{\sqrt{2}\beta{\bf B}_0}{\sqrt{2}\beta + c B_0} \,, \quad \delta{\bf H} = \frac{H_0}{B_0}\left(\delta{\bf B} - \frac{c {\bf B}_0 \,{\bf H}_0\cdot\delta{\bf B}}{\sqrt{2}\beta B_0}\right) \label{y4}
\end{equation}
and 
\begin{equation}
\frac{H_0}{B_0} = \frac{\sqrt{2}\beta}{\sqrt{2}\beta + c B_0} \,. \label{y5}
\end{equation}
The parameter $H_0/B_0 \leq 1$ plays a significant r\^ole. For instance, from Eq. (\ref{e29}), we have $\tilde{\omega}_p = (B_0/H_0)^{1/2}\omega_p$. 

Assume plane wave perturbations proportional to $\exp[i({\bf k}\cdot{\bf r} - \omega t)]$, for ${\bf B}_0 = B_0\hat{z}$, ${\bf k} = k (\sin\theta, 0, \cos\theta)$, so that $\theta$ is the angle between ${\bf k}$ and the equilibrium magnetic field. The linear wave analysis in this context is well-known \cite{Stix, Swanson, Bittencourt}. We shall isolate the linearized velocity field $\delta{\bf u} = (\delta u_x, \delta u_y, \delta u_z)$ from Eq. (\ref{y1}) in terms of $\delta{\bf E} = (\delta E_x, \delta E_y, \delta E_z)$, yielding as usual 
\begin{equation}
\delta u_x = \frac{e}{m}\frac{(\omega_c \delta E_y + i \omega \delta E_x)}{(\omega_c^2 - \omega^2)} \,, \quad \delta u_y = \frac{e}{m}\frac{(-\omega_c \delta E_x + i \omega \delta E_y)}{(\omega_c^2 - \omega^2)} \,, \quad \delta u_z = - \frac{i e \delta E_z}{m\omega} \,. \label{y6}
\end{equation}
Using ${\bf k}\times\delta{\bf E} = \omega\delta{\bf B}$ and inserting the results from Eq. (\ref{y6}) into the modified Amp\`ere-Maxwell law (\ref{y2}) yields a linear homogeneous system for the electric field components, 
\begin{equation}
\begin{bmatrix}
   {\cal S} - \eta^2\cos^{2}\theta &
   -i {\cal D} & \eta^2\cos\theta\sin\theta \\
   i {\cal D} & {\cal S} - \alpha(\theta)\eta^2 & 0 \\
	\eta^2\cos\theta\sin\theta & 0 & {\cal P} - \eta^2\sin^{2}\theta
   \end{bmatrix}
	\begin{bmatrix} \delta E_x \\ \delta E_y \\ \delta E_z \end{bmatrix} = 0 \,, \label{y7}
	\end{equation}
where Eqs. (\ref{y3}) and (\ref{y4}) were also needed. Here, $\eta = ck/\omega$ is the refraction index, $\alpha(\theta) = \cos^{2}\theta + (H_0/B_0)\sin^{2}\theta$  and 
\begin{equation}
{\cal D} = \frac{\omega_c \tilde\omega_p^2}{\omega(\omega_c^2-\omega^2)} \,, \quad {\cal S} = 1 + \frac{\tilde\omega_p^2}{\omega_c^2-\omega^2} \,, \quad {\cal P} = 1 - \frac{\tilde\omega_p^2}{\omega^2} \,.
\label{y8}
\end{equation}
are modified ${\cal D}$ (Difference), ${\cal S}$ (Sum) and ${\cal P}$ (Plasma) coefficients. 
In the limit $H_0/B_0 \rightarrow 1$, one recovers the usual result \cite{Stix, Swanson, Bittencourt}. 
Setting ${\cal P} = 0$ we get $\omega^2 = \tilde{\omega}_p^2$, while setting ${\cal S} = 0$ yields the modified upper-hybrid wave in Eq. (\ref{e28}). 

The determinant of the matrix in Eq. (\ref{y7}) must vanish. Following, as closely as possible the traditional notation \cite{Stix, Swanson, Bittencourt}, we find 
\begin{equation}
A \eta^4 - B \eta^2 + C = 0 \,, \label{y9}
\end{equation}
where 
\begin{eqnarray}
A &=& \alpha(\theta)\,({\cal S} \sin^{2}\theta + {\cal P} \cos^{2}\theta) \,, \\
B &=& {\cal R} {\cal L} \sin^{2}\theta + {\cal P} {\cal S}\, [\alpha(\theta) +  \cos^{2}\theta] \,,\\
C &=& {\cal P R L} \,, \quad {\cal R} = {\cal S} + {\cal D} \,, \quad {\cal L} = {\cal S} - {\cal D} \,,
\end{eqnarray}
also introducing modified R (Right) and L (Left) coefficients. 

After standard rearrangements,  
the solution to Eq. (\ref{y9}) can be expressed as 
\begin{equation}
\eta^2 = 1 - \,\frac{2(A-B+C)}{2A-B \pm \sqrt{B^2-4 A C}} \,,
\end{equation}
or 
\begin{equation}
\eta^2 = 1 - \frac{\tilde\omega_p^2/\omega^2}{Q} \,, \quad Q = Q_0 \pm F/Q_1 \label{y10}
\end{equation}
where 
\begin{eqnarray}
Q_0 &=& 1 + \frac{(1\!-\!2\alpha(\theta))\omega_c^2\tilde\omega_p^2\sin^{2}\theta+(\alpha(\theta)\!-\!1)\omega_c^2(\omega^2+\tilde\omega_p^2)-(\alpha(\theta)\!-\!1)(\omega^2-\tilde\omega_p^2)^2}{2\Bigl((\omega^2-\tilde\omega_p^2)(\tilde\omega_p^2+(\alpha(\theta)\!-\!1)\omega^2)-(\alpha(\theta)\!-\!1)\omega_c^2\omega^2\sin^{2}\theta\Bigr)} \,, \\
Q_1 &=& \frac{[(2\!-\!\alpha(\theta))\omega^2-\tilde\omega_p^2]\tilde\omega_p^2 + (\alpha(\theta)\!-\!1)\omega^4 - (\alpha(\theta)\!-\!1)\omega_c^2\omega^2\sin^{2}\theta}{(\omega^2-\omega_c^2)\omega^2/2} \,,\\
F^2 &=& [{\cal R}{\cal L}- (H_0/B_0){\cal P} {\cal S}]^2 \sin^{4}\theta + 4 \alpha(\theta) {\cal P}^2 {\cal D}^2 \cos^{2}\theta \,.
\end{eqnarray}

In the low field limit $H_0/B_0 \rightarrow 1$, one has 
\begin{equation}
Q = 1 - \frac{\omega_c^2 \sin^{2}\theta}{2(\omega^2-\omega_p^2)} \pm \left(\frac{\omega_c^4 \sin^{4}\theta}{4(\omega^2-\tilde\omega_p^2)^2} + \frac{\omega_c^2 \cos^{2}\theta}{\omega^2}\right)^{1/2} \,,
\end{equation}
which is the standard result \cite{Stix, Swanson, Bittencourt}.

Equation (\ref{y10}) is a generalized Hartree-Appleton equation, describing electromagnetic wave propagation in cold, uniform  plasma governed by a logarithmic electrodynamics. In the low field limit, the usual Appleton-Hartree equations \cite{AH} are recovered, in the collisionless case. Since ion motion was neglected, to apply Eq. (\ref{y10}) the frequency must be large in comparison with the ion cyclotron frequency.

Since the analysis of Eq. (\ref{y10}) is quite involved in general, we consider the special cases of propagation parallel or perpendicular to the ambient magnetic field.

\section{Propagation parallel to ${\bf B}_0$}

For $\theta = 0$, we have $\alpha(\theta) = 1$, irrespective of the strength of $B_0$. We also find, from Eq. (\ref{y10}), that $Q = 1 \pm \omega_c/\omega$, leading to a modified right-hand circularly polarized (RCP) wave,
\begin{equation}
\frac{c^2 k^2}{\omega^2} = 1 - \frac{\tilde\omega_p^2}{\omega(\omega-\omega_c)} 
\end{equation}
and a modified left-hand circularly polarized wave (LCP) wave, 
\begin{equation}
\frac{c^2 k^2}{\omega^2} = 1 - \frac{\tilde\omega_p^2}{\omega(\omega+\omega_c)} \,
\end{equation}
where the only change in comparison with the usual RCP and LCP waves is the replacement $\omega_p \rightarrow \tilde\omega_p$. 
The properties of RCP and LCP waves are well-known \cite{Stix, Swanson, Bittencourt} and are reproduced here provided the larger effective plasma frequency is used. 

However, the parallel propagation case admits another possibility, since for $\theta = 0$ all three coefficients $A, B, C$ in Eq. (\ref{y9}) become proportional to ${\cal P}$. Setting ${\cal P} = 0$, one regains the modified electron plasma oscillations $\omega^2 = \tilde\omega_p^2$. 

\section{Propagation perpendicular to ${\bf B}_0$}

For $\theta = 90^{\circ}$, one has $\alpha(\theta) = H_0/B_0$ and two wave modes, described below. 

\subsection{Modified ordinary mode}

Taking the minus sign in Eq. (\ref{y10}), we have $Q = 1$ and $\omega^2 = \tilde\omega_p^2 + c^2 k^2$, which is the ordinary (O) mode, modified by the presence of the new plasma frequency $\tilde{\omega}_p$, instead of the usual one. With this proviso, the standard analysis of the O-mode applies \cite{Stix, Swanson, Bittencourt}. 

\subsection{Modified extraordinary mode}

Taking the plus sign in Eq. (\ref{y10}), we are left with an involved expression of $Q$ and the dispersion relation
\begin{equation}
\frac{H_0}{B_0} \frac{c^2 k^2}{\omega^2} = \frac{(\omega^2-\tilde\omega_R^2)(\omega^2-\tilde\omega_L^2)}{\omega^2(\omega^2- \tilde\omega_h^2)} \,, \label{xm}
\end{equation}
where
\begin{equation}
\tilde\omega_R = \frac{1}{2}\left(\omega_c+(\omega_c^2+4\tilde\omega_p^2)^{1/2}\right) \,, \quad \tilde\omega_L = \frac{1}{2}\left(-\omega_c+(\omega_c^2+4\tilde\omega_p^2)^{1/2}\right) \,. \label{z1}
\end{equation}
and where $\tilde\omega_h$ is the modified upper-hybrid frequency defined in Eq. (\ref{e28}). 
Apart from the overall factor $H_0/B_0$ on the left-hand side of Eq. (\ref{xm}), we have the same result as the dispersion relation of the well-known extraordinary (X) mode, with the replacements $\omega_p \rightarrow \tilde\omega_p$, $\omega_h \rightarrow \tilde\omega_h$. 

\begin{figure}[ht]
\begin{center}
\includegraphics[width=4.5in]{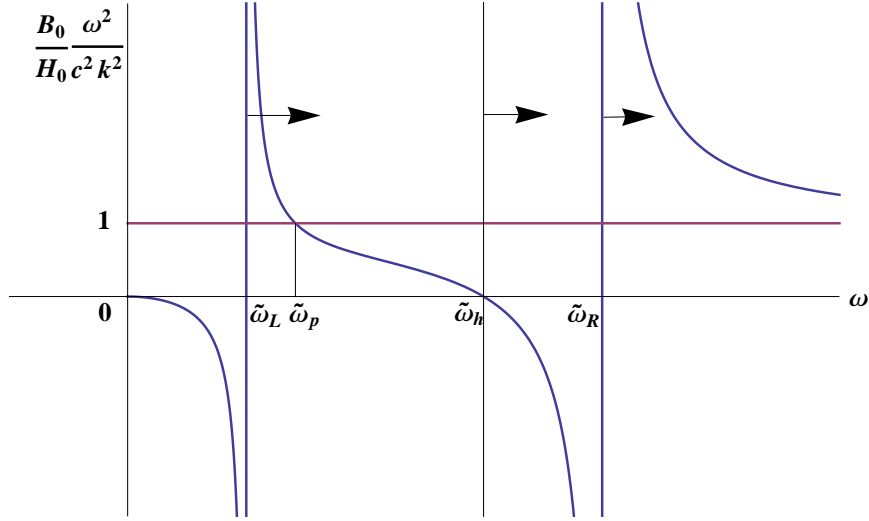}
\caption{Dispersion relation of the modified X-wave from Eq. (\ref{xm}). The arrows indicate the effects from the logarithmic electrodynamics. Forbidden bands: $0 < \omega < \tilde\omega_L$ and $\tilde\omega_h < \omega < \tilde\omega_R$.}
\label{fig1} 
\end{center}
\end{figure}

The analysis of the modified X-mode is more involved than for the other waves considered so far, but it may be performed with the aid of the definitions of cutoff and resonance. We recall \cite{Chen} that a cutoff happens whenever the refraction index, $\eta$, goes to zero, while a resonance occurs if $\eta$ becomes infinity. In general, a wave is absorbed at a resonance and reflected at a cutoff. From Eq. (\ref{xm}), we have that the location of cutoffs and resonances for the modified X-mode are the same as for the standard case, but in terms of the new plasma and upper-hybrid frequencies. Therefore, the standard analysis apply, with the net result shown in Fig. \ref{fig1}. Notice that the X-wave only propagates for $\tilde\omega_L < \omega < \tilde\omega_H$ or for $\omega > \tilde\omega_R$. 

It is apparent, from Eq. (\ref{z1}), that $\tilde\omega_L$ increases due to the logarithmic electrodynamic effects, with the corresponding increase of the left forbidden band in Fig. \ref{fig1}. In addition, for the allowed band $\tilde\omega_L < \omega < \tilde\omega_h$, one has
\begin{equation}
\tilde\omega_h - \tilde\omega_L = \omega_h - \omega_L - \Delta \omega_p^2 \left(\frac{1}{\sqrt{\omega_c^2+4\omega_p^2}}-\frac{1}{2\sqrt{\omega_c^2+\omega_p^2}}\right) + {\cal O}({\Delta^2}) \,, \quad \Delta = \frac{c B_0}{\sqrt{2}\beta} \,, \label{mn}
\end{equation}
while for the forbidden band  $\tilde\omega_h < \omega < \tilde\omega_R$  one has
\begin{equation}
\tilde\omega_R - \tilde\omega_h = \omega_R - \omega_h + \Delta \omega_p^2 \left(\frac{1}{\sqrt{\omega_c^2+4\omega_p^2}}-\frac{1}{2\sqrt{\omega_c^2+\omega_p^2}}\right) + {\cal O}({\Delta^2}) \,,   \label{m}
\end{equation}
where $\omega_{R,L}$ is the limit of $\tilde\omega_{R,L}$ in standard electrodynamics. The correction terms in Eq. (\ref{mn}) and (\ref{m}) are always negative (positive), showing a smaller allowed band and bigger forbidden band due to the log effects. The conclusion can be shown to be true for an arbitrary order too. 

In particular, for strongly magnetized plasmas, where $\omega_c \gg \omega_p$, one has, up to ${\cal O}(\Delta)$, 
\begin{equation}
\tilde\omega_h - \tilde\omega_L = \omega_c \left(1-\frac{\omega_p^2}{2\omega_c^2}\right) - \frac{\Delta\omega_p^2}{2\omega_c}  \,, \quad \tilde\omega_R - \tilde\omega_h = \frac{\omega_p^2}{2\omega_c}(1 + \Delta)  \,. \label{mm}
\end{equation}
For convenience, we recall that $\omega_p^2/\omega_c^2 = \mu_0 m n_0 c^2/B_{0}^2 = 1.02 \,n_0/B_{0}^2$, so that Eq. (\ref{mm}) requires $B_0 \gg \sqrt{n_0}$, always in terms of S. I. units. 

Similarly, for high density plasmas, such that $\omega_p \gg \omega_c$, one has
\begin{equation}
\tilde\omega_h - \tilde\omega_L = \omega_c \left(1+\frac{3\omega_c}{4\omega_p}\right) - \frac{3\Delta\omega_c^2}{16\omega_p}  \,, \quad \tilde\omega_R - \tilde\omega_h = \frac{\omega_c}{2}\left(1 -  \frac{3\omega_c}{4\omega_p}\right) + \frac{3\Delta\omega_c^2}{16\omega_p} \,,
\end{equation}
up to ${\cal O}(\Delta)$.

\section{Physical estimates}

We may present at least an estimate of the strength of the new effects, shown in the previous Sections, concerning wave propagation in plasma. We notice that the main changes come from the parameter 
\begin{equation}
\frac{\tilde\omega_p}{\omega_p} = \left(1 + \frac{c B_0}{\sqrt{2}\beta}\right)^{1/2} \,.
\end{equation}
If we estimate $\beta/c \sim 10^{11}\,{\rm T}$, we find $\beta \sim 10^{19} {\rm V/m}$, which is one order of magnitude larger than the Schwinger limit $m^2 c^3/(e\hbar) \sim 10^{18} {\rm V/m}$, where $\hbar$ is the reduced Planck's constant. Therefore, significant changes in plasma wave propagation may occur, for strongly magnetized plasma with $B_0 \sim 10^{11} \,{\rm T}$. Although the result has been obtained in the cold, uniform, collisionless and non-relativistic approximation, it is expected that the main conclusion remains true for more general plasmas, together with additional features. 

A possible candidate for such strongly magnetized plasmas would be the surface of a magnetar, with magnetic fields $B \sim 10^{10}-10^{12} \,{\rm T}$, where standard atomic nuclei compose a solid lattice together along with a sea of electrons \cite{magnetar}. However, in this case, more involved and detailed models would be necessary, with a more appropriate equation of state in connection with quantum and general- relativistic effects \cite{Shapiro}. Nevertheless, the relevance of the parameter $\tilde\omega_p/\omega_p$ will certainly show up also in a more detailed treatment. Ultra-strong magnetic fields beyond the QED limit are also found in other extreme astrophysical environments,  such as the interior of neutron stars, Central Engines of Supernovae and Gamma-Ray Bursts (GRB) and inner parts of GRB jets \cite{Uz}. Finally, the attainability of the Schwinger limit in laboratory with extreme power lasers should be mentioned \cite{Bu}.

\section{Concluding Comments and New Prospects}

In a first attempt to analyze plasma waves in a logarithmic electrodynamics, we have chosen a cold uniform plasma immersed in a magnetic field, not only because of the mathematical tractability, but also because a large number of wave modes in more detailed treatments may be associated with the modes obtained within the simplest approach. In the words by Stix \cite{Stix}, ``the 
cold-plasma model gives, in fact, a remarkably accurate description of the common small-amplitude perturbations that are possible for a hot plasma". 

Within the cold-plasma model, we have found a modified Trivelpiece-Gould dispersion relation for electrostatic waves in magnetized plasma, along with adequate changes of the plasma and upper-hybrid frequencies, due to the logarithmic electrodynamics effects. Allowing for electromagnetic waves yields a generalized Appleton-Hartree dispersion relation. The cases of propagation parallel or perpendicular to the equilibrium magnetic field are analyzed in detail, so that, in particular, generalized ordinary and extraordinary modes are found.  We determined the changes, due to logarithmic electrodynamics, in the allowable and forbidden frequency bands of the new extraordinary mode. 
According to physical estimates, non-linear electrodynamics effects become unavoidable for ultra-strong magnetic fields, which exist in extreme astrophysical plasma environments. Such developments are relevant for the determination of the $\beta$ parameter entering the basic new NLED Lagrangian density.

NLED models appear as effective photonic descriptions that take into account the sum over the quantum effects of virtual (charged) particle-antiparticle pairs. So, from the very onset, they are supposed to correctly describe electromagnetic effects associated to waves whose (wave)lengths are much bigger than the Compton wavelength of the charged particle whose quantum effects have been integrated over. In the case of electronic matter, we are talking about typical frequencies higher than $10^{20} $ Hz. So, our results for the logarithmic electrodynamics give a good description in this region of frequencies.

There is a number (around $15$) of NLED models in the literature. Among those, Born-Infeld electrodynamics is widely studied in many scenarios and, besides eliminating the singularity in the electric sector, it exhibits a more involved structure of electric and magnetic fields than the logarithmic model we have contemplated here, for it also involves the Lorentz-invariant quantity ${\bf E} \cdot {\bf B}$. So, as a further step, we wish to consider Born-Infeld model in a plasma medium, and this investigation would allow a new estimation of the Born-Infeld parameter, based on Plasma Physics. In the case of Born-Infeld and those models for which ${\bf E} \cdot {\bf B}$ is present, the ${\bf D}-$ and ${\bf H}-$fields are respectively modified by the additions of a term in ${\bf B}$ and ${\bf E}$. This brings about a new feature that is not contemplated in our present contribution. We shall be soon reporting on that elsewhere,  providing new insights about the existing connections \cite{Trines} between Born-Infeld theory and wave propagation at arbitrary angles, now in the context of a warm magnetized plasma.

{\bf Acknowledgments}: 
F.~H.~acknowledges the support by Con\-se\-lho Na\-cio\-nal de De\-sen\-vol\-vi\-men\-to Cien\-t\'{\i}\-fi\-co e Tec\-no\-l\'o\-gi\-co (CNPq). This study was financed in part by the Coordena\c{c}\~ao de Aperfei\c{c}oamento de Pessoal de N\'{\i}vel Superior - Brasil (CAPES) - Finance Code 001. LPRO is supported by the PCI-DB funds (CNPq/MCTIC). He expresses his gratitude to COSMO - CBPF for the friendly hospitality. P. G. was partially supported by Fondecyt (Chile) grant 1180178 and by Proyecto Basal FB0821. 

\end{document}